\documentclass[a4paper,twocolumn,english,prl,showpacs,superscriptaddress]{revtex4}
\pdfoutput=1
\usepackage{mathpazo}

\usepackage[T1]{fontenc}
\usepackage[latin9]{inputenc}
\usepackage{fancyhdr}
\usepackage{babel}
\usepackage{array}
\usepackage{amsmath}
\usepackage{amssymb}
\usepackage{graphicx}
\usepackage[unicode=true,pdfusetitle,
 bookmarks=true,bookmarksnumbered=false,bookmarksopen=false,
 breaklinks=false,pdfborder={0 0 1},backref=false,colorlinks=false]
 {hyperref}

\makeatletter

\pdfpageheight\paperheight
\pdfpagewidth\paperwidth

\providecommand{\tabularnewline}{\\}

\@ifundefined{textcolor}{}
{%
 \definecolor{BLACK}{gray}{0}
 \definecolor{WHITE}{gray}{1}
 \definecolor{RED}{rgb}{1,0,0}
 \definecolor{GREEN}{rgb}{0,1,0}
 \definecolor{BLUE}{rgb}{0,0,1}
 \definecolor{CYAN}{cmyk}{1,0,0,0}
 \definecolor{MAGENTA}{cmyk}{0,1,0,0}
 \definecolor{YELLOW}{cmyk}{0,0,1,0}
}

\usepackage{babel}
\usepackage{calrsfs}
\usepackage{hyperref}
\hypersetup{
    colorlinks=true,
    linkcolor=red,
    citecolor=blue,
    filecolor=magenta,      
    urlcolor=cyan,
}
\makeatother

\begin{document}

\title{Negative response with optical cavity and traveling wave fields }

\author{R. J. de Assis}

\address{Instituto de Física, Universidade Federal de Goiás, 74.001-970, Goiânia
- GO, Brazil}

\author{C. J. Villas-Boas}

\address{Departamento de Física, Universidade Federal de São Carlos, 13565-905,
São Carlos, São Paulo, Brazil}

\author{N. G. de Almeida}

\address{Instituto de Física, Universidade Federal de Goiás, 74.001-970, Goiânia
- GO, Brazil}

\pacs{05.30.-d, 05.20.-y, 05.70.Ln}
\begin{abstract}
We present a feasible protocol using traveling wave field to experimentally
observe negative response, \emph{i.e}., to obtain a decrease in the
output field intensity when the input field intensity is increased.
Our protocol uses one beam splitter and two mirrors to direct the
traveling wave field into a lossy cavity in which there is a three-level
atom in a lambda configuration. In our scheme, the input field impinges
on a beam splitter and, while the transmitted part is used to drive
the cavity mode, the reflected part is used as the control field to
obtain negative response of the output field. We show that the greater
cooperativity of the atom-cavity system, the more pronounced the negative
response. The system we are proposing can be used to protect devices
sensitive to intense fields, since the intensity of the output field,
which should be directed to the device to be protected, is diminished
when the intensity of the input field increases.
\end{abstract}
\maketitle

\section{\textup{Introduction}\textit{ }}

Negative response is a counter-intuitive effect, in the sense that
given an input, the output behaves contrary to what is expected. As
for example, cooling by heating, meaning the possibility of slowing
down the motion of a given system by increasing the temperature of
its reservoir, was proposed for optomechanical implementation \cite{Mari12},
solid state device \cite{Cleuren12}, and radiation-matter interaction
between a two-level atom and a single mode of a traveling wave field
in the trapped ion domain \cite{norton12}. Recently, a connection
between non-equilibrium thermal correlations and negative response
as given by cooling by heating was investigated for a system composed
by two atoms interacting with a single electromagnetic mode of a lossy
cavity \cite{Almeida16}.

In this paper, we investigate negative response for electromagnetic
wave field intensity using a single mode of an optical field and a
three-level atom in a lambda configuration inside a cavity. In our
proposal, a traveling wave field whose strength is $\varepsilon_{in}=\varepsilon$
enters a \emph{blackbox} and an output field of amplitude $\varepsilon_{out}$
leaves this blackbox, such that increasing the input field $\varepsilon$,
the output field $\varepsilon_{out}$ is reduced, or, conversely,
decreasing $\varepsilon$, the output field $\varepsilon_{out}$ is
raised. We investigate the parameter regimes which optimize this effect,
taking into account the main dissipative channels. Our device, working
in the negative response regime, has potential application in security
systems such as electronic devices sensitive to abrupt changes of
the input field intensity. For instance, by placing our device in
front of field detectors which is enough sensitive to count one or
two photons would avoid potential damage caused by input of undesired
intense fields. As an effective application, consider the relevant
problem of experimentally fake violation of Bell's inequality by exploiting
the physics of single photon detectors \cite{Macarov09,Gerhardt11,Scarani11}.
As Bell inequality is used, among others, to certificate nonlocal
channels as well as to guarantee secure quantum communication \cite{Acin06,Bancal 11},
it is important to prevent it from fake violations. In Ref. \cite{Scarani11},
the authors take advantage of the detailed working mechanism of the
avalanche single photon detectors, which becomes blind beyond some
optical power level, to manipulate their outputs in a controlled manner,
conveniently simulating the arriving of a single photon in any detector
they want. In this way, they induce photocounts in the detectors they
choose, leading to arbitrary violations of the Bell inequality, including
unphysical violations greater than the limit allowed by quantum mechanics
\cite{Popescu94}. Thus, it becomes desirable some device preventing
this problem in an automated way, without needing to monitor the input
optical power \cite{Scarani11}. In this regard, a device based on
the negative response principle developed here could circumvent this
problem, since any attempt to saturate the photodetectors by using
high intensities would not work, since the input intensity would be
reduced to that close to a single photon.

\section{Model}

\begin{figure}[ptbh]
\centering{}\includegraphics[bb=25bp 0bp 624bp 369bp,scale=0.425]{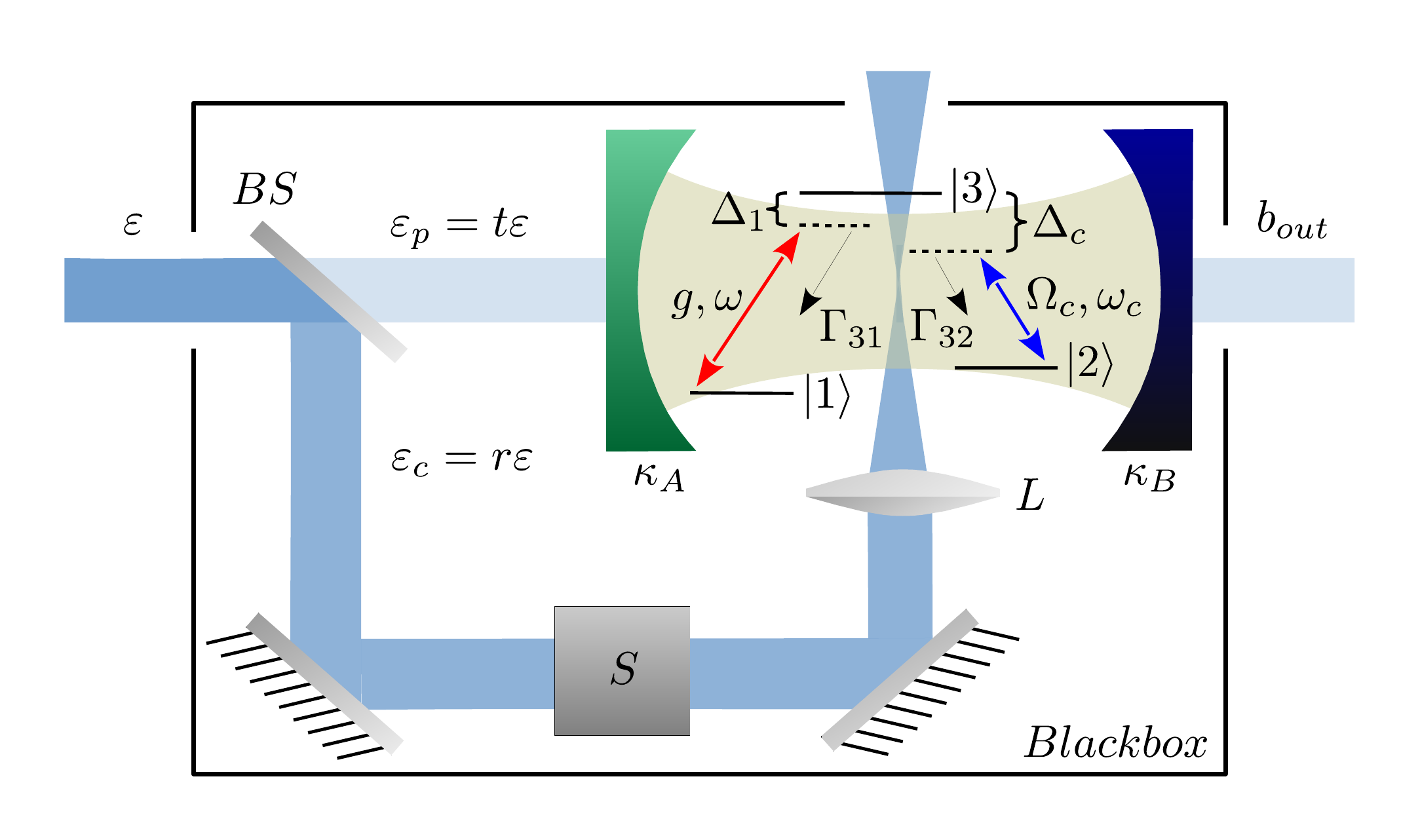}\caption{\label{fig:1}Pictorial representation of the experimental setup.
An input field of strength $\varepsilon$ impinges on a beam splitter
whose transmission and reflection coefficients are $t$ and $r$,
respectively, and the reflected and transmitted beam are directed
to a three-level atom inside an optical cavity. The transmitted part
is used as a driving field $\varepsilon_{P}$ on the cavity mode,
while the reflected part is used as the control field $\varepsilon_{c}$.
A polarizer $S$ and a lens $L$ are placed in the way of the control
field $\varepsilon_{c}$ to adjust its polarization and the Rabi frequency
to properly interact with a specific atomic transition. The other
parameters that appear in this figure are described in the main text.
By increasing $\varepsilon$ we simultaneously increase the driving
$\left(\varepsilon_{p}\right)$ and control $\left(\varepsilon_{c}\right)$
fields.}
\end{figure}

In Fig. \ref{fig:1} we represent pictorially the physical system
corresponding to our protocol to implement negative response with
optical fields and cavity. In our proposal a driving field of strength
$\varepsilon$, which serves as input, enters a blackbox and is either
transmitted or reflected by a beam splitter $BS$ whose transmission
and reflection coefficients are $t$ and $r$, respectively. The transmitted
beam $\varepsilon_{p}=t\varepsilon$, which is used as the driving
field, is sent through the left wall of a lossy  optical cavity of
damping $\kappa_{A}$, and the reflected beam $\varepsilon_{c}=r\varepsilon,$
which is used as the control field, is directed to the open side of
the optical cavity, both interacting with a three-level atom in a
lambda configuration. The output field emerges from the blackbox from
the right wall whose damping is $\kappa_{B}$. Note that increasing
the input field $\varepsilon$ we simultaneously increase both the
control $\left(\varepsilon_{c}\right)$ and the driving $\left(\varepsilon_{p}\right)$
fields. The relation between the outside and intracavity modes are
given by \cite{collett1984} 
\[
a_{out}=\sqrt{2\kappa_{A}}a-\varepsilon_{P},
\]
\[
b_{out}=\sqrt{2\kappa_{B}}a,
\]
being $a_{out}$ and $b_{out}$ the annihilation operators for the
outside fields, left and right modes, respectively, while $a$ is
the annihilation operator for the intracavity mode. Thus, once the
dynamics of the intracavity field is known, the transmitted beam can
straightforwardly be obtained. 

In this system a single three-level atom in a $\Lambda$ configuration
has two ground states $\left\vert 1\right\rangle $ and $\left\vert 2\right\rangle $
and excited state $\left\vert 3\right\rangle $. This three-level
atom interacts with one mode of the optical cavity of frequency $\omega$
inducing $\left\vert 1\right\rangle \leftrightarrow\left\vert 3\right\rangle $
transitions with Rabi frequency $g$. In turn, the control field of
frequency $\omega_{c}$ induces $\left\vert 2\right\rangle \leftrightarrow\left\vert 3\right\rangle $
transitions with Rabi\textbf{ }frequency given by $\Omega_{c}=\frac{\mu\varepsilon_{c}}{\hbar}=r\mu\sqrt{\frac{\omega_{c}\left\langle n\right\rangle }{2\hbar\epsilon_{0}V_{c}}}$,
where $V_{c}$ is the quantization volume regarded to the control
field and can be diminished using a lens to focus the beam, $\epsilon_{0}$
is the vacuum permittivity, $\mu$ is the atomic dipole transition
matrix element, and the average number of photons $\left\langle n\right\rangle $
of the input field. The Hamiltonian corresponding to the driving field
onto the cavity mode impinging the optical cavity through the left
wall is given by $\varOmega_{p}ae^{i\omega_{p}t}+h.c.$, where $h.c.$
stands for Hermitian conjugate, and $\varOmega_{p}=-i\sqrt{2\kappa_{A}}\varepsilon_{P}$
is the Rabi frequency for the probe field. Adopting state $\left\vert 1\right\rangle $
as the zero energy level, the total Hamiltonian $H=H_{0}+H_{int}$
reads $\left(\hbar=1\right)$ 
\[
H_{0}=\omega_{3}\sigma_{33}+\omega_{2}\sigma_{22}+\omega a^{\dagger}a,
\]
\[
H_{int}=ga\sigma_{31}+\Omega_{c}\sigma_{32}e^{-i\omega_{c}t}+\Omega_{p}ae^{i\omega_{p}t}+h.c.,
\]
\[
\sigma_{ij}=\left\vert i\right\rangle \left\langle j\right\vert ,i,j=1,2,3.
\]
In the interaction picture we can write this Hamiltonian as 
\begin{multline}
H_{I}(t)=ga\sigma_{31}e^{i\Delta_{1}t}+\Omega_{c}\sigma_{32}e^{i\Delta_{c}t}+\Omega_{p}ae^{i\Delta_{p}t}+h.c,\label{eq:1}
\end{multline}
where $\Delta_{1}\equiv\omega_{3}-\omega$, $\Delta_{c}\equiv\left(\omega_{3}-\omega_{2}\right)-\omega_{c}$,
and $\Delta_{p}=\omega_{p}-\omega$ . The time-dependency of the interaction
Hamiltonian Eq. \eqref{eq:1} can be eliminated applying the unitary
transformation $U_{1}=e^{-i\left[\Delta_{p}a^{\dagger}a-\Delta_{1}\sigma_{33}-\left(\Delta_{1}-\Delta_{c}\right)\sigma_{22}-\Delta_{p}\sigma_{11}\right]t}$,
which allows us to rewrite this Hamiltonian as 
\begin{multline}
\widetilde{H}_{I}=\Delta_{1}\sigma_{33}+(\Delta_{1}-\Delta_{c})\sigma_{22}+\Delta_{p}\sigma_{11}-\\
-\Delta_{p}a^{\dagger}a+\left[\Omega_{p}a+ga\sigma_{31}+\Omega_{c}\sigma_{32}+h.c.\right].\label{eq:2}
\end{multline}

\begin{figure}[h]
\centering{}\includegraphics[bb=-3bp 0bp 576bp 432bp,scale=0.46]{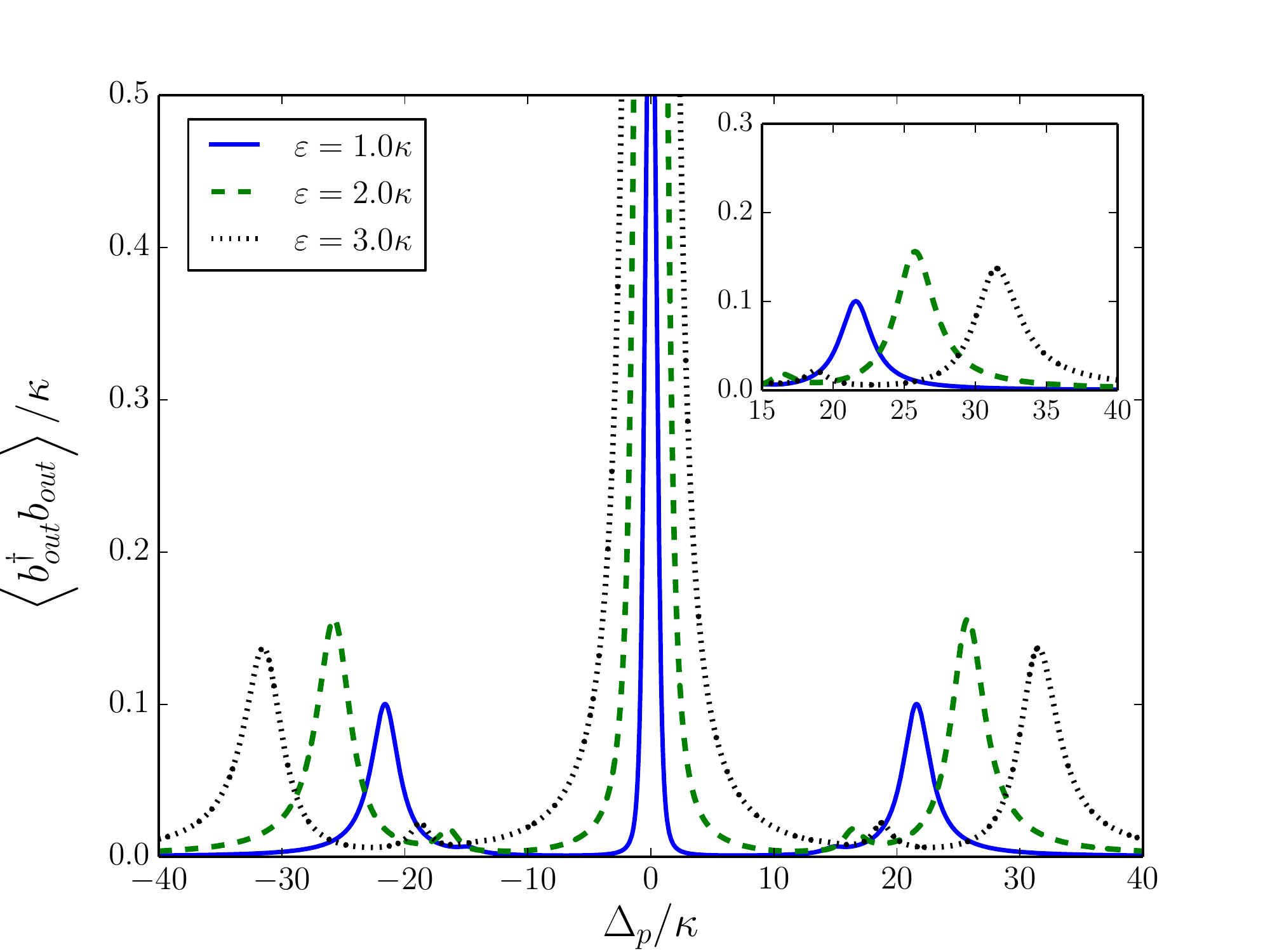}\caption{\label{fig:2}Average photon number transmission \emph{versus} the
normalized detuning between the probe and the cavity mode field ($\Delta_{P}/\kappa$)
for different values of the driving field strength $\varepsilon$.
The solid (blue) line is for $\varepsilon=1.0\kappa$, dashed (green)
line is for $\varepsilon=2.0\kappa$, dotted (black) line is for $\varepsilon=3.0\kappa$.
Here $g=20\kappa$ and $\Gamma=\kappa$.}
\end{figure}

\begin{figure}[h]
\begin{centering}
\begin{tabular}{>{\centering}m{0.3\columnwidth}>{\centering}m{0.7\columnwidth}}
\includegraphics[bb=8bp 0bp 269bp 340bp,scale=0.305]{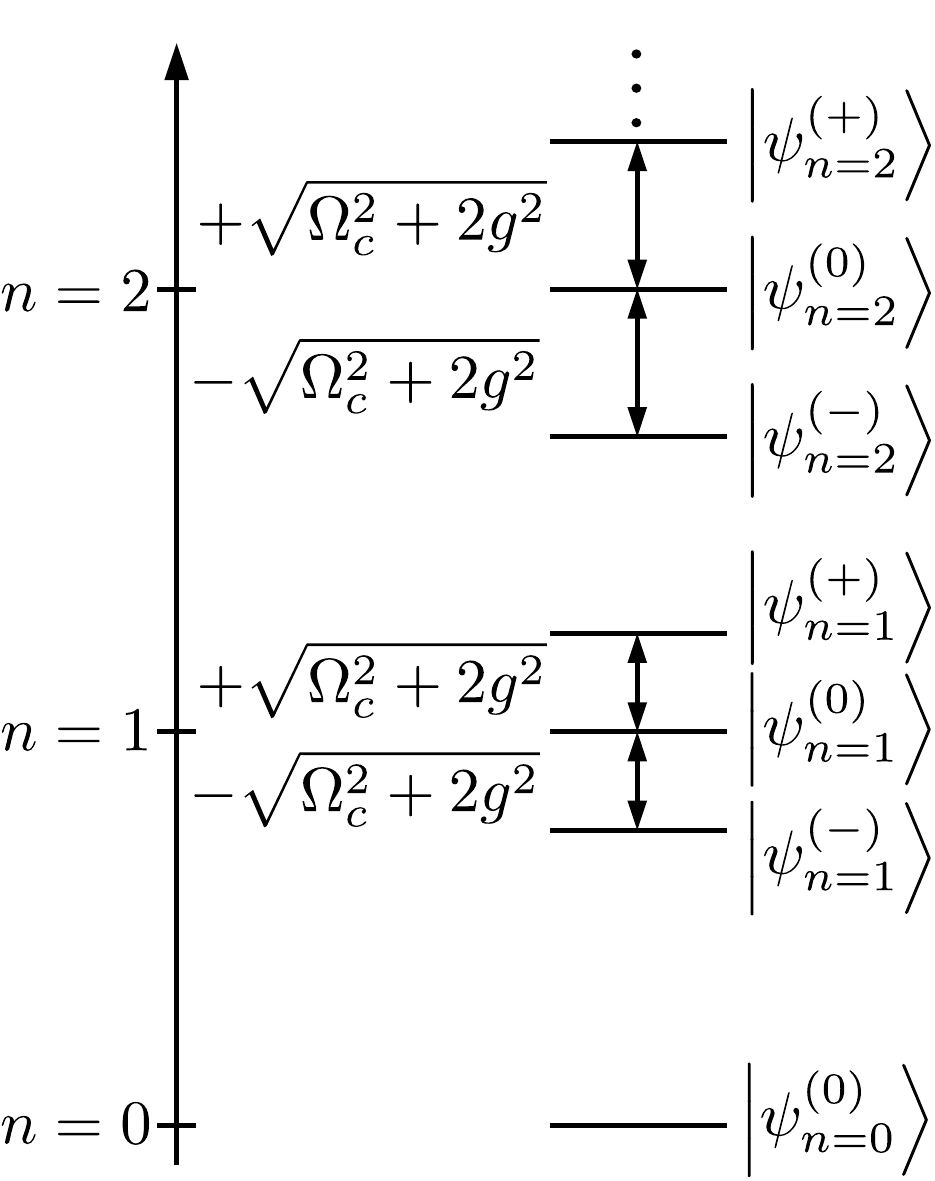}  & \includegraphics[bb=-6bp 0bp 576bp 432bp,scale=0.31]{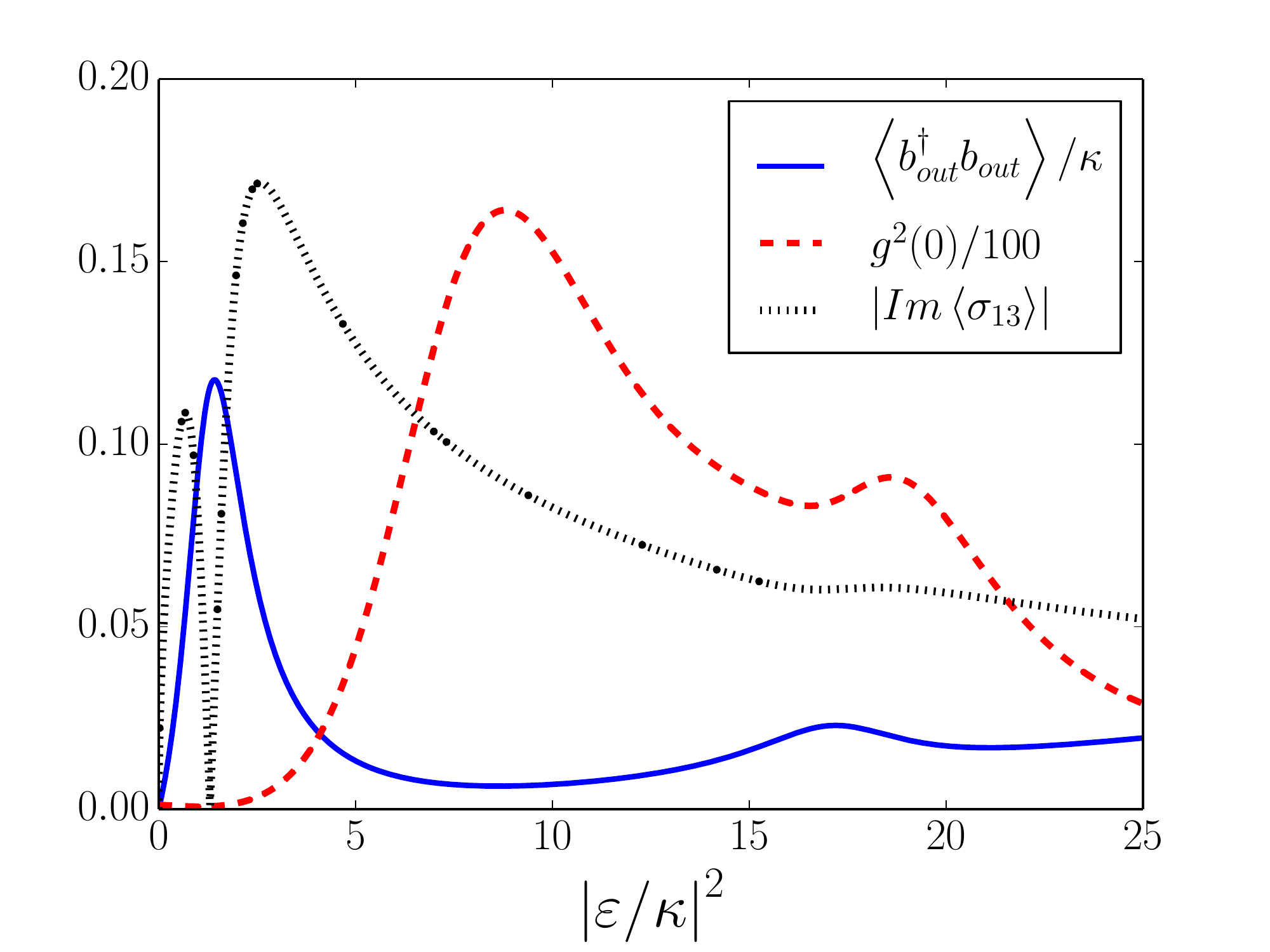}\tabularnewline
(a)  & (b)\tabularnewline
\end{tabular}
\par\end{centering}
\caption{\label{fig:3}(a) Firsts eigenstates $\left({\scriptstyle \left|\psi_{n}^{\left(0\right)}\right\rangle },{\scriptstyle \left|\psi_{n}^{\left(\pm\right)}\right\rangle },n=0,1,2,...\right)$
of the Hamiltonian system given by Eq. \eqref{eq:2}, considering
all detunings and $\varOmega_{p}$ nulls \cite{Celso13}. From this
figure we can see that adjusting the Rabi frequency of control field
$\Omega_{c}$ we can tune one or two (or even higher) photon resonances,
as explained in the text. (b) Average photon number of the output
field $\left\langle b_{out}^{\dagger}b_{out}\right\rangle /\kappa$
(blue solid line), correlation function $g^{(2)}\left(0\right)$,
divided by $100$ for clarity (red dashed line), and the atomic absorption
$\left|\text{\text{Im}}\left\langle \sigma_{13}\right\rangle \right|$
(black dotted line) as a function of the normalized input average
photon number $\left|\varepsilon/\kappa\right|^{2}$. }
\end{figure}

The master equation corresponding to Eq. \eqref{eq:2} is 
\begin{multline}
\frac{d\rho}{dt}=-i[\widetilde{H}_{I},\rho]+\kappa D\left[a\right]+\Gamma_{31}D\left[\sigma_{13}\right]+\\
+\Gamma_{32}D\left[\sigma_{23}\right]+\gamma_{2}D\left[\sigma_{22}\right]+\gamma_{3}D\left[\sigma_{33}\right],\label{master}
\end{multline}
where $\Gamma_{ij}$ is the atomic polarization decay rate from the
state $\left\vert i\right\rangle \rightarrow\left\vert j\right\rangle $,
$\gamma_{i}$ is the atomic dephasing of the state $\left\vert i\right\rangle $,
$\kappa$ stands for the total cavity field decay rate, i.e., $\kappa=\kappa_{A}+\kappa_{B}$,
and $D\left[A\right]=2A\rho A^{\dagger}-A^{\dagger}A\rho-\rho A^{\dagger}A$.
We investigate the negative response of this system by solving the
above master equation in the steady state regime. As this equation
does not present analytical solution for arbitrary values of the parameters,
we\textbf{ }numerically solve it by using QuTip algorithms \cite{Qutip}.
In our simulations we are assuming a symmetric cavity, i.e., $\kappa_{A}=\kappa_{B}=\kappa/2$,
$\kappa$ being our reference parameter and $\Gamma_{31}=\Gamma_{32}=\Gamma/2$.
Since the dephasing rates of the levels $\left\vert 2\right\rangle $
and $\left\vert 3\right\rangle $ ($\gamma_{2}$ and $\gamma_{3}$)
are usually very small as compared to the damping rates, we neglect
them throughout this paper. Before presenting our main results, we
stress that the Hamiltonian model above leads to the well known phenomena
of electromagnetic induced transparency \cite{celso2010,Celso13},
as shown in Fig. \ref{fig:2}, where the average photon number outgoing
the cavity $\left\langle b_{out}^{\dagger}b_{out}\right\rangle /\kappa$
\emph{versus} $\Delta_{p}/\kappa$ is shown. In this figure the solid
(blue) line is for $\varepsilon=1.0\kappa$, the dashed (green) line
is for $\varepsilon=2.0\kappa$, and the dotted (black) line is for
$\varepsilon=3.0\kappa$. Note the central peak denoting the maximum
of the transmission, which occurs for $\omega_{p}=\omega$. Also,
note the secondary peaks, whose maxima are far apart from the center
$\omega_{p}=\omega$ by $\pm\sqrt{ng^{2}+\Omega_{c}^{2}}$ where $n=1,2,...$
is the photon number of the eigenstates of the Hamiltonian given by
Eq. \eqref{eq:2}, obtained considering all detunings and $\varOmega_{p}$
nulls \cite{Celso13}. These eigenstates, schematically shown in Fig.
\ref{fig:3} (a), were explicitly derived in Ref. \cite{Celso13}.
In Fig. \ref{fig:3} (b), the second order correlation function $g^{(2)}\left(0\right)$
(red dashed line), here divided by $100$ for convenience, the atomic
absorption $\left|\mathbf{\text{Im}}\left\langle \sigma_{13}\right\rangle \right|$
(black dotted line), and the average photon number of the output field
(solid blue line) are plotted as a function of the normalized average
photon number $\left|\varepsilon/\kappa\right|^{2}$ of the input
field. The parameters used were $\Gamma_{31}=\Gamma_{32}=\Gamma/2=\kappa/2$,
$\Delta_{1}=\Delta_{c}=0$, $\Delta_{p}=1.1g$, $\Omega_{c}=8\varepsilon$,
and $g=20\kappa$. Note that exactly on the maximum of the first peak,
as seen from left to right, corresponding to the average photon number
of the output state (solid blue curve), we see that the correlation
function $g^{(2)}(0)$ (dashed red line) goes down close to zero,
which indicates the predominance of the one-photon process. A similar
behavior occurs for the second peak, where the maximum of the average
photon number of the output practically coincides with a local minimum
of the $g^{(2)}(0)$ function, indicating now a large contribution
of the two-photon process. 

Turning back to Fig. \ref{fig:2} and its inset, note that by increasing
the strength of the driving field $\varepsilon$ the lateral peaks
move away as both the driving $\left(\varepsilon_{P}\right)$ and
the control $\left(\varepsilon_{c}\right)$ fields onto the cavity
mode and the atom, respectively, depend directly on $\varepsilon$.
As a consequence of this movement by the lateral peaks, the following
effect can be seen: driving the system with a fixed detuning, for
example $\Delta_{p}\simeq22\kappa$, which is the detuning providing
the maximum transmission for the blue solid line $\left(\varepsilon=1.0\kappa\right)$,
we see that for larger values of the driving field $\left(\varepsilon\right)$,
the transmission for this specific detuning goes down as the lateral
peak moves to the right, as seen from the dash (green) and dot (black)
curves at the same point.
\begin{figure}[h]
\centering{}\includegraphics[bb=-3bp 0bp 576bp 432bp,scale=0.46]{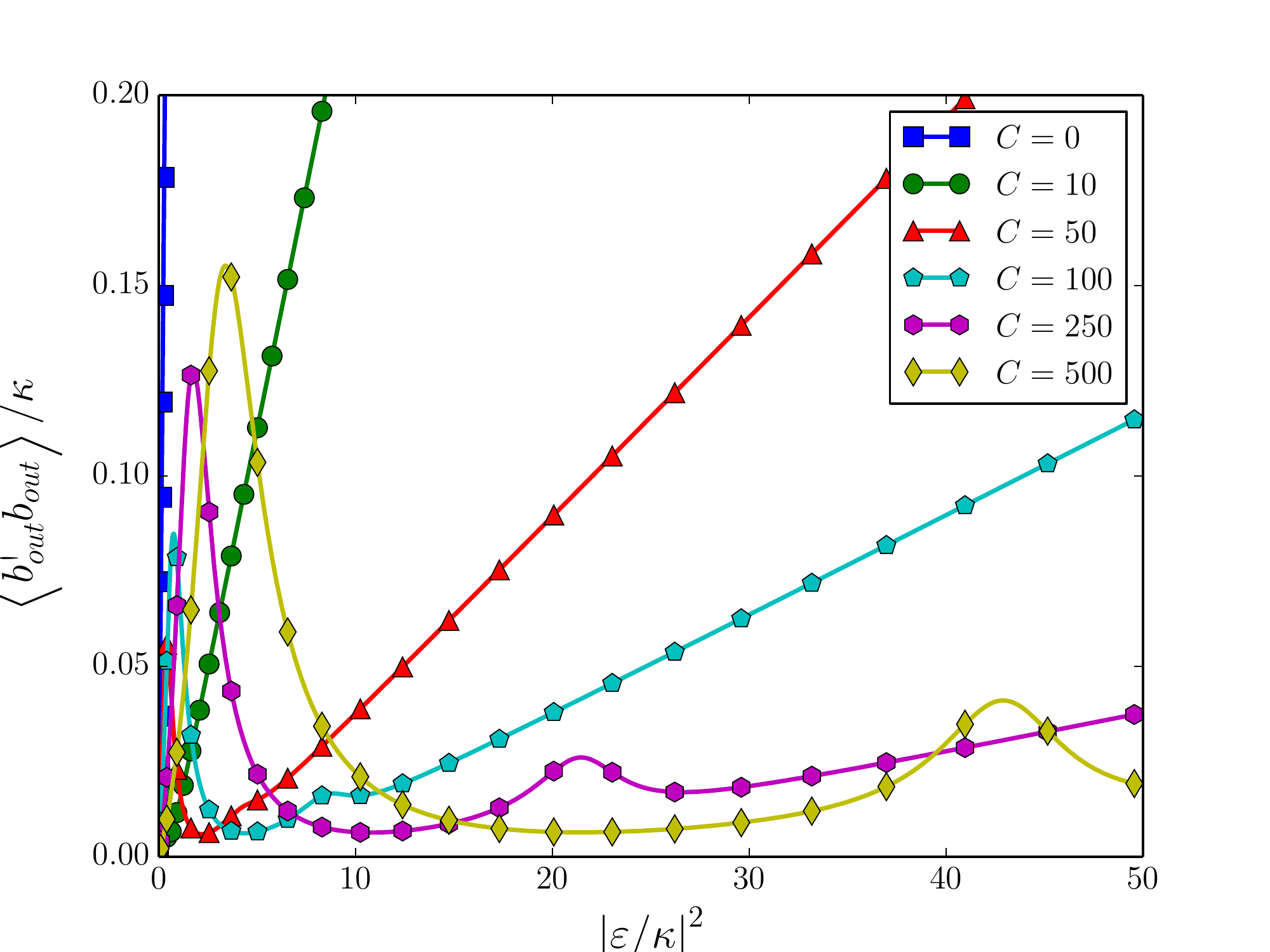}\caption{\label{fig:4}Average photon number of the output field \emph{versus}
normalized input field, $\left|\varepsilon/\kappa\right|^{2}$ for
i) $C=0$ (square blue line); ii) $C=10$ (circle green line); iii)
$C=50$ (triangle red line); iv) $C=100$ (pentagon aqua line); $C=250$
(hexagon violet line), and $C=500$ (diamond golden line). Here we
used the following parameters: $\Gamma_{31}=\Gamma_{32}=\Gamma/2=\kappa/2$,
$\Delta_{1}=\Delta_{c}=0$, $\Delta_{p}=1.1g$, and $\Omega_{c}=8\varepsilon$.\vspace{-0.25cm}
}
\end{figure}

\section{Results }

Now we investigate in details the negative response in this system.
To be clear, we are going to increase the input field strength $\varepsilon$,
thus simultaneously increasing both the control $\varepsilon_{c}$
and the driving $\varepsilon_{P}$ cavity fields, according to Fig.
\ref{fig:1}, in order to obtain a reduction in the output beam intensity.

Our main result is shown in Fig. \ref{fig:4}, where we plot the output
photon number average $\left\langle b_{out}^{\dagger}b_{out}\right\rangle /\kappa$
\emph{versus} the rescaled input field intensity $\left|\varepsilon/\kappa\right|^{2}$
for several values of the cooperativity $C=g^{2}/(2\kappa\varGamma)$.
Here we note that although we have used $t/r=4$, this ratio is not
important to accomplish our proposal, since the control field Rabi
frequency $\varOmega_{c}$ can be adjusted by simply focusing the
reflected beam, once $\Omega_{c}$ is inversely proportional to the
square root of the volume $V_{c}$ of the beam interacting with the
atom. The parameters used here were $\Gamma_{31}=\Gamma_{32}=\Gamma/2=\kappa/2$,
$\Delta_{1}=\Delta_{c}=0$, $\Delta_{p}=1.1g$, $\Omega_{c}=8\varepsilon$,
$\varOmega_{p}=-0.8i\sqrt{\kappa_{A}}\varepsilon$, and $C$ ranges
from $0$ to $500$. To these parameters we can see two peaks in Fig.
\ref{fig:4}, depending on the values of $C$, with the negative response
starting at the maximum and ending at the subsequent minimum. As for
example, for $C=250$ (hexagon violet line), negative response start
in the first peak, as seen from the left to the right, around $\left|\varepsilon/\kappa\right|^{2}=2$,
and finishes at $\left|\varepsilon/\kappa\right|^{2}=10$, starting
again in the second peak, around $\left|\varepsilon/\kappa\right|^{2}=22$,
finishing at $\left|\varepsilon/\kappa\right|^{2}=27$. Therefore,
starting at the first peak as seen from the left, we see the transmission
going down as we increase the strength of the driving field due to
the movement to the right of the main peak. Then, the transmission
increases again, since the system reaches a two-photon resonance,
and subsequently it goes down. Finally, the transmission begins to
increase to larger values of the input field as the atomic system
saturates. For very large values of the input field, the average photon
number inside the cavity increases linearly with $\left|\varepsilon/\kappa\right|^{2}$,
whose behavior is identical to that of an empty cavity coherently
driven (not shown in Fig. \ref{fig:4}).

\begin{figure}[ptbh]
\centering{}\includegraphics[bb=10bp 0bp 576bp 432bp,scale=0.46]{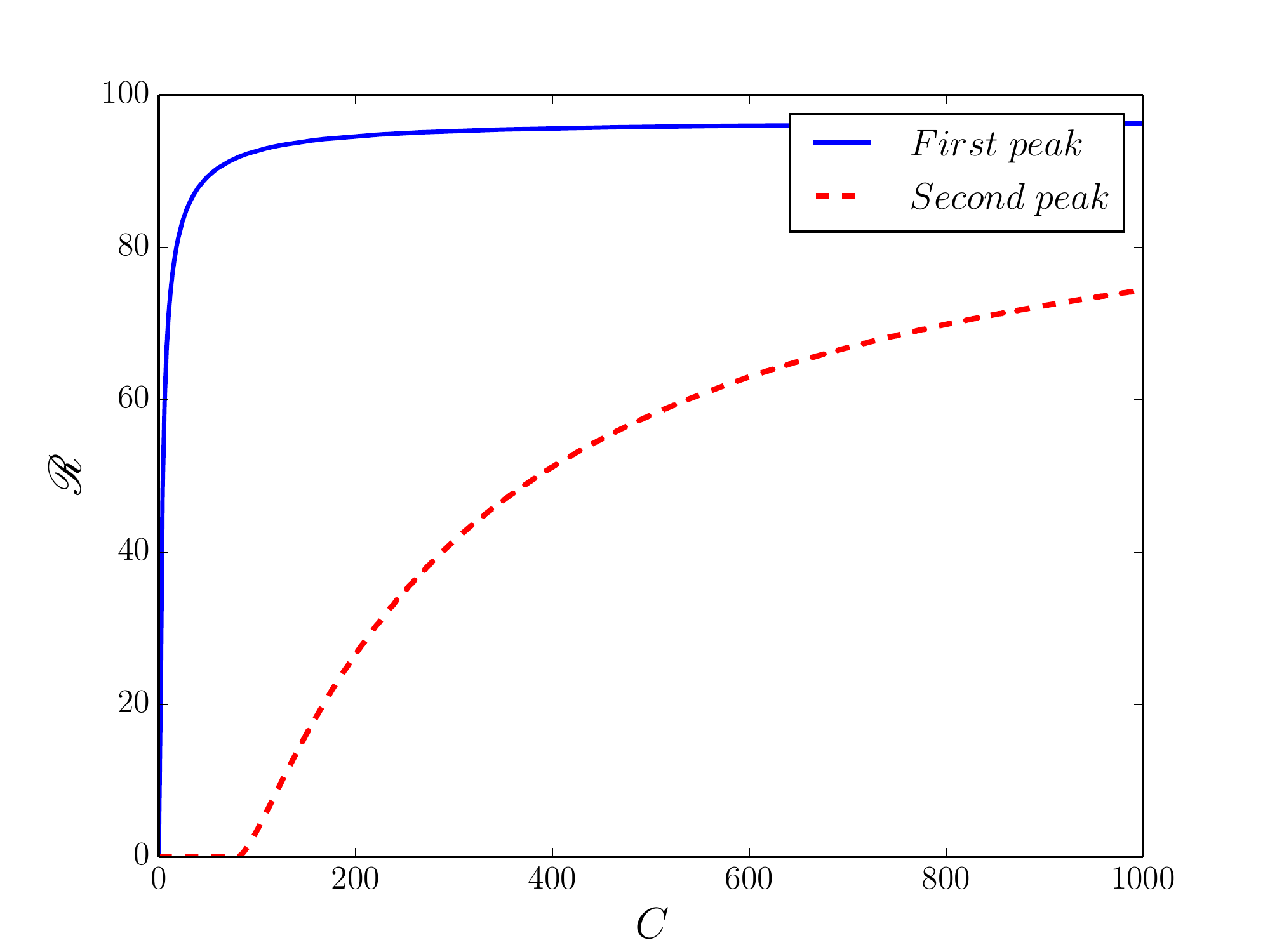}\caption{\label{fig:5}The parameter $\mathcal{R}$, a measure of negative
response, gives the percentage of the output beam variation as a function
of the cooperativity $C$. Here we plot $\mathcal{R}$ for both the
one-photon resonance (first peak in solid blue line), and the two-photon
resonance (second peak in red dashed line). Note that the maximum
variation for both one- and -two photon resonances stabilize near
of 95\% (solid-blue line) and 80\% (dashed-red line). Here, first
peak (second peak) refers to the leftmost (rightmost) peak appearing
in Fig. \ref{fig:4}.\vspace{-0.5cm}
}
\end{figure}

Since each cooperativity $C$ determines different negative responses,
it is useful to define a parameter $\mathcal{R}$ that can be used
in order to quantify the negative response in terms of the variation
in the output beam intensity. Indeed, cooperativity $C$ giving rise
to a very small negative response could be not attractive from the
experimental point of view. We thus propose the parameter $\mathcal{R}=\frac{\left(\left\langle b_{out}^{\dagger}b_{out}\right\rangle _{max}-\left\langle b_{out}^{\dagger}b_{out}\right\rangle _{min}\right)}{\left\langle b_{out}^{\dagger}b_{out}\right\rangle _{max}}\times100\%$
relating the maximum value taken by the output beam in the first (second)
peak and the subsequent minimum right after the first (second) peak.
In Fig. \ref{fig:5} we show this parameter, which gives the percentage
of the output beam variation, as a function of the cooperativity $C$,
for both the first (solid, blue line) and the second (dashed, red
line) peaks. Note that the larger the parameter $\mathcal{R}$, the
higher the percentage of the output beam variation, allowing the experimentalist
to choose $C$ conveniently to guarantee that the negative response
be evaluated or even optimized. Note that, to the parameters used
here, $\mathcal{R}$ increases monotonically with the cooperativity,
reaching the maximum around 95\% for the one-photon resonance (solid-blue
line) and around 80\% for two-photon resonance (dashed-red line).
Also, note that negative response for one-photon resonance (solid
blue line), as measured by the $\mathcal{R}$ parameter, initiates
at small values of the cooperativity $\left(C>0\right)$ and grows
faster up to $C\simeq60$, saturating after $C\sim200$, while for
two-photon resonances (dashed-red line), $\mathcal{\mathcal{R}}$
initiates at $C\simeq85$, growing steadily and saturating for $C\sim1000$.
Interesting, to this case note that there is no negative response
for cooperativity values lesser than $C\simeq85$, no matter the intensity
of the input field. This is due to the difficulty of having two-photon
processes for small values of the cooperativity.

\section{\textit{\emph{Conclusions}} }

In this paper we have studied negative response in the context of
optical cavity and traveling wave field. We presented an experimentally
feasible scheme to observe decreasing of the output field intensity
while increasing the intensity of the input field, therefore a negative
response phenomenon. We characterize this negative response for a
large range of atom-quantum field Rabi frequencies, and we were able
to propose the parameter $\mathcal{R}$ that quantifies the efficiency
of this effect through the percentage of the negative variation of
the output field when the input field is positively varied. We also
showed that negative response can be displayed to either one-photon
and two-photon resonances, with one-photon resonance requiring lower
values of the cooperativity. Among some applications, we pointed out
that our proposal can be helpful to protect fragile devices against
sudden variation of field intensities, such as field detectors sensitive
to few photons. In particular, a device made with the principles developed
here could be used to defeat the fake Bell violation strategy used
elsewhere \cite{Scarani11}, which consists in impinging strong field
intensity to blind single photon detectors, manipulating their outputs
to simulate the arriving of a single photon in the detector that they
choose conveniently.
\begin{acknowledgments}
We acknowledge financial support from the Brazilian agency CNPq, CAPES
and FAPEG. This work was performed as part of the Brazilian National
Institute of Science and Technology (INCT) for Quantum Information.
C.J.V.-B. acknowledges support from Brazilian agencies No. 2013/04162-5
Sao Paulo Research Foundation (FAPESP) and from CNPq (Grant No. 308860/2015-2). 
\end{acknowledgments}

\end{document}